\documentclass[article]{IEEEtran}
\usepackage[latin1]{inputenc}
\usepackage[T1]{fontenc}
\usepackage{times}
\usepackage{cite}
\usepackage{pifont}
\usepackage{url}
\usepackage{tikz}
\usepackage{circuitikz}
\usepackage{amsmath}
\usepackage{listings}
 \usetikzlibrary{angles}
 \usetikzlibrary{positioning}
 \usetikzlibrary{decorations}
\usepackage{algorithm}
\usepackage{algpseudocode}
\usepackage{setspace}
\usetikzlibrary{decorations.pathmorphing}
\usetikzlibrary{shapes.symbols}
\usepackage{tikz-3dplot}
\usepackage{amssymb}
\newcommand{\M}[1]{\mathbf{#1}}
\newcommand{\C}[1]{\hat{\mathbf{#1}}}
\newcommand{\T}[1]{\mathrm{#1}}
\newcommand{\V}[1]{\boldsymbol{#1}}
\renewcommand{\u}[1]{\boldsymbol{\hat{#1}}}

\usetikzlibrary{shapes.geometric}
\usetikzlibrary{decorations.pathreplacing}

\newcommand{\ie}{\textit{i}.\textit{e}.}
\newcommand{\eg}{\textit{e}.\textit{g}.}

\newcommand{\kv}{\boldsymbol{\hat{k}}}
\newcommand{\rv}{\V{r}}

\title{Equivalent External Noise Temperature of Time-Varying Receivers}

\author{Kurt Schab, \IEEEmembership{Member, IEEE}, and K.C. Kerby-Patel, \IEEEmembership{Senior Member, IEEE}
\thanks{Manuscript received \today; revised \today. \textit{(Corresponding author: Kurt Schab.)}}
\thanks{K. Schab is with Santa Clara University, Santa Clara, USA (e-mail: kschab@scu.edu).}
\thanks{K.C. Kerby Patel is with the University of Massachussetts, Boston, USA (e-mail: kc.kerby-patel@umb.edu).}
}

\begin{document}

\setcounter{figure}{0}
\setcounter{section}{0}
\newpage
\pagestyle{headings}
\twocolumn

\maketitle

\begin{abstract}
    The equivalent external noise temperature of time-varying antennas is studied using the concept of cross-frequency effective aperture, which quantifies the intermodulation conversion of external noise across the frequency spectrum into a receiver's operational bandwidth.  The theoretical tools for this approach are laid out following the classical method for describing external noise temperature of linear time-invariant antennas, with generalizations made along the way to capture the effects of time-varying components or materials.  The results demonstrate the specific ways that a time-varying system's noise characteristics are dependent on its cross-frequency effective aperture and the broadband noise environment.  The general theory is applied to several examples, including abstract models of hypothetical systems, antennas integrated with parametric amplification, and time-modulated arrays.  
\end{abstract}

\begin{IEEEkeywords}
Antenna theory, noise, time-varying circuits.
\end{IEEEkeywords}

\section{Introduction}

\IEEEPARstart{T}{ime-varying} systems and devices couple signals of dissimilar temporal frequencies, leading to mixing, up-conversion, and down-conversion.  
Recently, there has been a growing interest in the use of periodic time-varying components within otherwise time-invariant antenna systems to achieve performance advantages such as amplification \cite{frost1960correspondence,ram1994parametric, loghmannia2019active, mekawy2021parametric} or impedance bandwidth enhancement \cite{manteghi2016wideband,singletary2021plasma, loghmannia2022broadband,slevin2022broadband}.  
More exotic effects, such as non-reciprocity \cite{wang2018time, chen2021efficient} and broadband perfect absorption \cite{pacheco2020antireflection,Li2021temporal,firestein2022absorption,wang2022broadband}, can also be realized in electromagnetic and acoustic systems of finite size using time-varying materials.  See~\cite{hayran2023using} for an extensive bibliography.   

The introduction of time-varying elements into a receiving system, however, naturally leads to out-of-band noise and undesired signals mixed into the observation bandwidth, potentially affecting the overall system signal-to-noise ratio (SNR).
This effect is well known in other areas of RF engineering. A similar process gives rise to intermodulation interference, or intermodulation distortion, in receiver design, in which strong signals in undesired bands mix in the nonlinear elements of the receiver and can appear in the output band of interest \cite{ITU-intermodulation},\cite[\S 5.2]{maas1993microwave},\cite[\S 13.5]{pozar2011microwave}. Induced intermodulation distortion has also been used to detect unknown radios at a distance \cite{kane2015detecting}.
In the same way, noise outside the desired RF input band can be unintentionally mixed into the output band of mixers when it appears at the IF frequency or in the other sideband and is not properly filtered before mixing \cite{hull1993systematic},\cite[\S 9.5.1]{maas2005noise}. Even desired signals can produce interference effects at the output if they overlap with the IF bandwidth, particularly in parametric amplifying receivers and receiving antennas that operate in or near the degenerate mode \cite[\S 6.3]{blackwell1961}, \cite{blosser2024signal}.

In this paper, we consider the effect of general time-variation  on the equivalent external noise temperature of a receiving antenna.   We adapt the classical approach to antenna noise temperature to the case of a time-varying antenna using a generalized cross-frequency formulation of effective aperture.  This generalized effective aperture approach is compatible with full-wave time-varying electromagnetic analysis via conversion-matrix / harmonic balance methods  \cite{bass2022conversion}, frequency-domain co-simulation \cite[\S 6.1]{maas2005noise} and covariance-based noise models \cite{russer2015, kerbypatel2023}, abstract models of time-modulated arrays, and transient simulations.
Assumptions regarding signal incidence and external noise are introduced throughout the text, with emphasis placed on following those used in the standard derivation of antenna noise temperature. 

Sections \ref{sec:models}-\ref{sec:tv} outline the general theory of the paper, with results summarized in Sec.~\ref{sec:summary}.  Sections~\ref{sec:demo}-\ref{sec:tma} describe examples selected from a variety of applications, with judicious simplifications applied to facilitate replication of critical results. Details on computational methods used throughout the paper, as well as extended derivations of specific results, are included in several appendices.

\section{Signal and Noise Models}
\label{sec:models}

\begin{figure}
        \centering
        \begin{tikzpicture}[scale=1.4,photon/.style={decorate,decoration={snake,post length=2mm}}]
            \draw[dashed] (0,0) circle (1);
            \def\theta{45}
            \draw[-stealth,photon,segment length=10pt,thick,blue] ({1.8* cos(\theta)},{1.8*sin(\theta)}) node[above left] {$\tilde{s}$} -- ({1.1* cos(\theta)},{1.1*sin(\theta)});

            \foreach \theta in {120,140,...,180}{
            \draw[-stealth,photon,segment length=10pt,thick,red] ({1.8* cos(\theta)},{1.8*sin(\theta)}) node[left] {$\tilde{n}$} -- ({1.1* cos(\theta)},{1.1*sin(\theta)});}
            \node[circle,fill=yellow!10,draw = black] at (0,0) {$\T{rx}$}; 
            
        \end{tikzpicture}
        \caption{General analysis setup. 
 Noise intensity $\tilde{n}$ impinges upon a receiver from all directions $\kv$, while a signal $\tilde{s}$ is incident from only the $\kv_\T{s}$ direction.}
        \label{fig:signal-models}
    \end{figure}
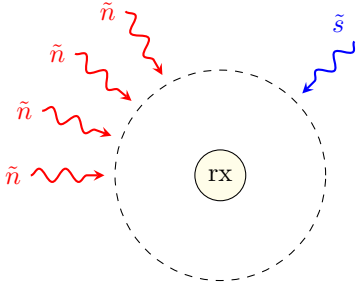

The setup for all subsequent analysis is shown in Fig.~\ref{fig:signal-models}, where all signals incident upon a receiver propagate as plane waves.  The noise intensity $\tilde{n}$ is incident upon a receiver from all directions $\kv$.  In general, the noise intensity may be angularly dependent~\cite{ITU-noise}, but here we follow a standard first-order simplification~\cite[\S 2.18]{balanis2016antenna} and assume an isotropic thermal noise corresponding to the frequency-dependent background temperature $T_\T{b}(\omega)$~\cite{sarabandi2022bandwidth},
\begin{equation}
\tilde{n}(\omega) = \frac{k_\T{B}}{\lambda^2}T_\T{b}(\omega).
\label{eq:noise-model}
\end{equation}
with units $\T{W}/(\T{m}^2\cdot\T{Hz}\cdot \T{sr})$.  Here $k_\T{B} = 1.38\cdot10^{-23}~\T{J}/\T{K}$ is Boltzmann's constant and $\lambda$ is the freespace wavelength associated with the frequency $\omega$.  Generalization to anisotropic noise enviroments is described in later sections.

In MIMO or multipath environments, an intended signal may arrive from multiple incident angles, however here we simplify analysis and assume that the intended signal is incident from a single direction $\kv_\T{s}$ and carries the power spectral density $s_0$.  Utilizing this assumption and matching the units of noise intensity, the signal intensity can be written as
    \begin{equation}
        \tilde{s}(\omega,\kv) = s_0(\omega)\delta(\kv-\kv_\T{s}).
        \label{eq:signal-model}
    \end{equation}
 The incident signal is assumed to be bandlimited to a bandwidth $B$ centered around a carrier frequency $\omega_0$, \ie,
    \begin{equation}
        s_0(\omega) = 0, \quad \omega\notin\Omega_\T{B},
    \end{equation}
    with
    \begin{equation}
        \Omega_\T{B} = \left[\omega_0 - B/2, \omega_0 + B/2\right],
    \end{equation}
    and the total power in the incident signal is fixed to
    \begin{equation}
        \int s_0(\omega) \T{d}\omega = S_0.
    \end{equation}
    The observation bandwidth of the receiver is selected to match the incident signal bandwidth, leading to a received system power of the form
    \begin{equation}
        P_\T{rx} = \int_{\Omega_\T{B}} p_\T{rx}(\omega) \T{d}\omega,
    \end{equation}
    where $p_\T{rx}$ is the power spectral density (signal, noise, or combined) observed at the receiver.

\section{LTI Receiver Analysis}
\label{sec:lti}

Consider a plane wave of frequency $\omega$ incident from the direction $\kv$ carrying power flux spectral density $y$ with units of $\T{W/(m^2\cdot Hz)}$.  The power spectral density delivered to a termination attached to a receiving antenna is determined by the antenna's realized effective aperture $A_\T{eff}(\omega,\kv)$ (including mismatch losses)
\begin{equation}
    p_\T{rx}(\omega) = A_\T{eff}(\omega,\kv) y(\omega,\kv).
    \label{eq:s-simple-source}
\end{equation}
The above expression can be elevated to account for uncorrelated sources distributed over the far-field sphere via integration\footnote{Here we adopt the convention that all integrations taken over the far-field sphere implicitly include a sum over two orthogonal polarizations.  Omitting this sum from all expressions greatly simplifies the notation and has no impact on the conclusions or results.}
\begin{equation}
    p_\T{rx}(\omega) = \int A_\T{eff}(\omega,\kv) \tilde{y}(\omega,\kv)\T{d}\kv,
    \label{eq:s-distributed-source}
\end{equation}
where $\tilde{y}$ is a power flux density per unit solid angle, i.e., intensity of the form of either \eqref{eq:noise-model} or \eqref{eq:signal-model}.  

\subsection{LTI received signal}
Consider a narrowband incident signal described by the assumptions in Sec.~\ref{sec:models} and the incident signal direction $\kv_\T{s}$.  The received signal power spectral density is then 
\begin{equation}
    p_\T{rx}^\T{s}(\omega) =  A_\T{eff}(\omega,\kv_\T{s})s_0(\omega),
\end{equation}
following the form of \eqref{eq:s-simple-source}.  If the signal bandwidth $B$ is narrow enough that the effective aperture is approximately constant and the load includes filtering to only receive power over that bandwidth, the total received power reduces to the classical result
\begin{equation}
    P_\T{rx}^\T{s} = A_\T{eff}(\omega_0,\kv_\T{s})S_0.
\end{equation}

\subsection{LTI received noise}
To calculate the corresponding noise power spectral density and total received noise power, we assume the general model for isotropic incident noise introduced in \eqref{eq:noise-model}. Assuming noise sources originating from different angles are uncorrelated, by \eqref{eq:s-distributed-source}, the total noise power spectral density at the receiving load is
\begin{equation}
    p_\T{rx}^\T{n} = \frac{k_\T{B}}{\lambda_0^2}T_\T{b}\int A_\T{eff}(\omega,\kv) \T{d}\kv.
    \label{eq:n-distributed-source}
\end{equation}
Assuming the effective aperture and noise characteristics are uniform over the receiver bandwidth, the total noise power is
\begin{equation}
    P_\T{rx}^\T{n} = \frac{k_\T{B}BT_\T{b}}{\lambda_0^2} \int  A_\T{eff}(\omega_0,\kv) \T{d}\kv.
    \label{eq:lti-noise-1}
\end{equation}
The integral can be evaluated using the reciprocal identity relating the radiation intensity $U$ and the total radiated power $P_\T{rad}$~\cite[\S 1.8]{stutzman2012antenna} when the system is operated in transmit mode
\begin{equation}
A_\T{eff}(\omega_0,\kv) = \frac{\lambda_0^2 G}{4\pi} = \frac{\lambda_0^2 \eta\tau U(\kv)}{P_\T{rad}},
\label{eq:aeff-g}
\end{equation}
which, when integrated over the far-field sphere gives
\begin{equation}
    \int A_\T{eff}(\omega_0,\kv) \T{d}\kv = \eta\tau\lambda_0^2.
    \label{eq:aeff-bar}
\end{equation}
Here $\tau$ and $\eta$ are the antenna's mismatch loss and radiation efficiency at the center frequency.  Using this result in \eqref{eq:lti-noise-1} leads to the standard definition of the external noise temperature of a receiving antenna~\cite{best2013realized}
\begin{equation}
    P_\T{rx}^\T{n} =  k_\T{B}B\left(\eta\tau T_\T{b}\right) = k_\T{B}BT_\T{a}^\T{LTI}
    \label{eq:lti-noise}
\end{equation}

The total signal-to-noise ratio under these assumptions is
\begin{equation}
    \T{SNR} = \frac{S_0A_\T{eff}(\omega_0,\kv_\T{s})}{k_\T{B}B\left(\eta\tau T_\T{b}\right)}.
    \label{eq:snr-lti}
\end{equation}
Note that losses on the antenna itself will contribute additional noise not considered here.  The exclusion of thermal noise due to losses on the antenna body is only valid when~\cite[Eq. 4]{best2013realized}
\begin{equation}
    \eta T_\T{b} \gg (1-\eta)T_0,
\end{equation}
with $T_0$ being the physical temperature of the antenna, often taken to be approximately $290-300~\T{K}$ for terrestrial applications~\cite[\S2.18]{balanis2016antenna}\cite[\S10.1]{pozar2011microwave}.

\begin{figure}
    \centering
        \begin{tikzpicture}[scale=0.9,transform shape]
    \draw[-stealth] (-0.5,0) -- (6.5,0) node[above] {$\omega$};
    \draw[ultra thick,red!80!black] (-0.5,1) -- (6,2);
    \draw[fill=red!80!black,fill opacity = 0.2,draw=none] (-0.5,0) -- (-0.5,1) -- (6,2) -- (6,0) -- (-0.5,0);
    \draw[-stealth] (-0.5,0) -- node[above,rotate=90,align=center] {external noise\\$\tilde{n}(\omega,\kv)$} (-0.5,2.5);
    \draw[dashed] (2.5,0) -- (2.5,2);
    \draw[dashed] (1.5,0) -- (1.5,2);
    \draw[dashed] (1.5,2) -- (2.5,2);
    \draw[stealth-stealth] (1.5,2.1) -- node[above] {$B$} (2.5,2.1);

    \begin{scope}[shift={(0,-4)}]
    \draw[-stealth] (-0.5,0) -- (6.5,0) node[above] {$\omega$};
    \draw[ultra thick,blue!80!black] (-0.5,1) -- (6,2);
    \draw[fill=blue!80!black,fill opacity = 0.2,draw=none] (-0.5,0) -- (-0.5,1) -- (6,2) -- (6,0) -- (-0.5,0);
    \draw[-stealth] (-0.5,0) --node[above,rotate=90,align=center] {received noise\\$p^\T{n}_\T{rx}(\omega)$} (-0.5,2.5);
    \draw[dashed] (2.5,0) -- (2.5,2);
    \draw[dashed] (1.5,0) -- (1.5,2);
    \draw[dashed] (2.5,2) -- (1.5,2);
    \draw (2,0)--(2,-0.2) node[below] {$\omega_0$};
    \draw (0.5,0)--(0.5,-0.2) node[below] {$\omega_{-1}$};
    \draw (3.5,0)--(3.5,-0.2) node[below] {$\omega_1$};
    \draw (5,0)--(5,-0.2) node[below] {$\omega_2$};
    \end{scope}

    \path[-stealth] (0.5,0.5) edge[bend left=20] node[left] {$\bar{A}_\T{eff}^{-1}$}(1.9,-3.5);
    \path[-stealth] (2,0.5) edge[bend left=20] node[above right] {$\bar{A}_\T{eff}^0$}(2,-3.5);
    \path[-stealth] (3.5,0.5) edge[bend left=20] node[above right] {~$\bar{A}_\T{eff}^1$}(2.1,-3.5);
    \path[-stealth] (5,0.5) edge[bend left=20] node[above right] {~~$\bar{A}_\T{eff}^2$}(2.2,-3.5);

    \draw[dashed] (0,0) -- (0,2);
    \draw[dashed] (1,0) -- (1,2);
    \draw[dashed] (0,2) -- (1,2);
    
    \draw[dashed] (3,0) -- (3,2);
    \draw[dashed] (4,0) -- (4,2);
    \draw[dashed] (3,2) -- (4,2);

    \draw[dashed] (4.5,0) -- (4.5,2);
    \draw[dashed] (5.5,0) -- (5.5,2);
    \draw[dashed] (4.5,2) -- (5.5,2);
    
    \end{tikzpicture}
    \caption{Depiction of an antenna's cross-frequency effective aperture coupling incoming noise from each modulation harmonic into the observation bandwidth.  Assuming a isotropic noise environment, the average terms $\bar{A}_\T{eff}^p$ control the amount of coupling from each harmonic via~\eqref{eq:tv-littlepnoise}.}
    \label{fig:out-of-band-schem}
\end{figure}
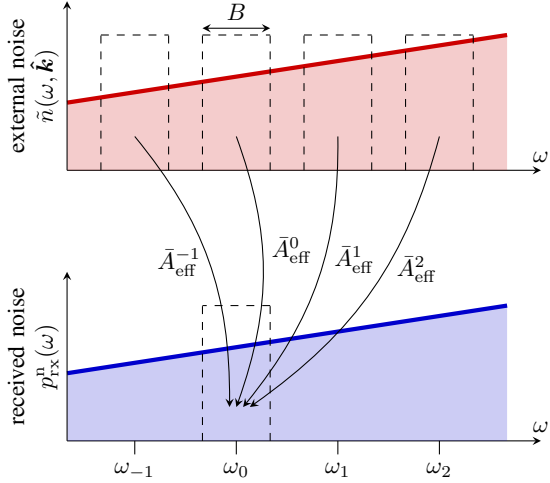

\section{Time-varying Receiver Analysis}
\label{sec:tv}

In a general time-varying system, incident signals at one frequency $\omega'$ give rise to induced currents, scattered fields, and received power at other frequencies $\omega$. The time-varying system response may also depend on the phase $\phi$ (or equivalently, a time offset) of the incident signal relative to the system's time variation, \eg, in systems based on degenerate-mode parametric amplification \cite{mekawy2021parametric, blosser2024signal}. This implies that the effective aperture now requires four arguments and the received power spectral density becomes
\begin{equation}
    p_\T{rx}(\omega,\phi) = \iint a_\T{eff}(\omega,\omega',\phi,\kv) \tilde{y}(\omega',\phi,\kv) \T{d}\kv \T{d}\omega'.
    \label{eq:def-cross-aeff}
\end{equation}
Here $a_\T{eff}(\omega,\omega',\phi,\kv)$ has units of $\T{m}^2/\T{Hz}$ and the interpretation of mapping power spectral density at frequency $\omega'$ with phase $\phi$ incident from the direction $\kv$ to the received power spectral density at frequency $\omega$.

Practical receiver systems do not \textit{a priori} know the phase of incident noise or desired signals.
Therefore, the phase of incident noise and signals must usually be treated as a uniformly distributed random variable.
We are typically interested in time-average power quantities, and since the instantaneous phase of noise or signals at any observation frequency is itself a random process, we compute the expected received power averaged over incident phase angle.  
In the remainder of this paper, we refer to powers and effective apertures averaged over uniformly-distributed incident phase.

For periodically time-varying systems, the effective aperture at the observation frequency $\omega$ is non-zero only for $\omega' = \omega+p\omega_\T{m}$, where $p$ is an integer and $\omega_\T{m}$ is the modulation frequency of the system~\cite[\S 3.4]{maas2003nonlinear}.  This reduces \eqref{eq:def-cross-aeff} to
\begin{equation}
    p_\T{rx}(\omega) = \sum_p \int A^p_\T{eff}(\omega,\kv) \tilde{y}(\omega+p\omega_\T{m},\kv)\T{d}\kv,
    \label{eq:aeffp-sum-form}
\end{equation}
where
\begin{equation}
    A^p_\T{eff}(\omega,\kv) = A_\T{eff}(\omega,\omega+p\omega_\T{m},\kv)
\end{equation}
is the periodic cross-frequency effective aperture mapping the $p^\T{th}$ harmonic frequency $\omega+p\omega_\T{m}$ onto the observation frequency $\omega$.  Note that this special case can be transformed back to the form of \eqref{eq:def-cross-aeff} by sythesizing the effective aperture $a_\T{eff}$ as a train of delta functions at $\omega' = \omega+p\omega_\T{m}$ weighted by the corresponding values of $A^p_\T{eff}$.

\subsection{TV received signal}
Unless an incident signal is particularly designed to have additional harmonic components mixed down to the receiver's matched observation bandwidth, the signal bandwidth would likely be selected to lie completely within the interval $\left[\omega - \omega_\T{m}/2,\omega + \omega_\T{m}/2\right]$, \ie, $B\leq \omega_\T{m}$, such that
\begin{equation}
    \tilde{s}(\omega+p\omega_m,\kv) = 0,\quad p \neq 0.
\end{equation}
In that case, the received signal power spectral density once again reduces to
\begin{equation}
    p_\T{rx}^\T{s}(\omega) = A^0_\T{eff}(\omega,\kv_\T{s}) s_0(\omega)
\end{equation}
and the total signal power, assuming relatively constant effective aperture over the receive bandwidth, is
\begin{equation}
    P_\T{rx}^\T{s} = A^0_\T{eff}(\omega_0,\kv_\T{s}) S_0.
\end{equation}
Note that the effective aperture $A^0_\T{eff}(\omega_0,\kv_\T{s})$ maps incident signals at (or near) frequency $\omega_0$ to the observation bandwidth.  Thus it has equivalent interpretation (though it may have a different numerical value) to the LTI effective aperture at this frequency and incident angle.

\subsection{TV received noise}
Unlike the intended signal, the external noise is not generally bandlimited and out-of-band noise can be mixed into the observation bandwidth.  Inserting the isotropic noise of the form of \eqref{eq:noise-model} into the model described by \eqref{eq:aeffp-sum-form}, the noise power spectral density within the receive bandwidth is
\begin{equation}
    p^\T{n}_\T{rx}(\omega) = \sum_p \frac{k_\T{B}T_\T{b}^p}{\lambda_p^2} \int A^p_\T{eff}(\omega,\kv) \T{d}\kv,
    \label{eq:prxn-tv-intermediate}
\end{equation}
with
\begin{equation}
    \lambda_p = \frac{2\pi c}{\omega+p\omega_\T{m}},
\end{equation}
where $c$ is the speed of light in vacuum and $T_\T{b}^p$ is the external brightness temperature at frequency $\omega+p\omega_\T{m}$.  Reciprocity may not hold for a general time-varying antenna \cite{anderson1965reciprocity}, so it may not be possible to reduce the integral on the above right-hand side using the relation between gain and effective aperture.  Nevertheless, the integral can be rewritten as the scaled average of $A_\T{eff}^p(\omega,\kv)$ over the far-field sphere,
\begin{equation}
    \bar{A}_\T{eff}^p(\omega) = \frac{1}{4\pi}\int A_\T{eff}^p(\omega,\kv)\T{d}\kv,
    \label{eq:aeffpbar-def}
\end{equation}
reducing the above expression to
\begin{equation}
    p_\T{rx}^\T{n}(\omega) = \sum_p\frac{4\pi k_\T{B} T_\T{b}^p \bar{A}^p_\T{eff}(\omega)}{\lambda_p^2}.
    \label{eq:tv-littlepnoise}
\end{equation}
This explicitly shows how cross-frequency effective aperture terms $\bar{A}_\T{eff}^p$ map out-of-band noise, quantified by the external noise intensity or brightness temperature, into the observation bandwidth, as depicted in Fig.~\ref{fig:out-of-band-schem}.  Invoking the assumption of relatively stationary quantities over the observation bandwidth, the received power is given by
\begin{equation}
    P_\T{rx}^\T{n} = k_\T{B} B \left(4\pi \sum_p\frac{T_\T{b}^p \bar{A}^p_\T{eff}(\omega_0)}{\lambda_p^2}\right) = k_\T{B}BT_\T{A}^\T{TV}
    \label{eq:tv-bigpnoise}
\end{equation}
The term $T_\T{A}^\T{TV}$ in the above expression is the effective external noise temperature observed by the time-varying antenna accounting for intermodulation effects. Note that in the LTI case, only the $p = 0$ term is non-zero and the average reciprocal effective aperture is equal to $\eta\tau\lambda^2_0/(4\pi)$, reducing the expression to match that in \eqref{eq:lti-noise}.  Under all of the preceding assumptions, the signal-to-noise ratio observed by the time-varying antenna is 
\begin{equation}
    \T{SNR} = \frac{A^0_\T{eff}(\omega_0,\kv_\T{s}) S_0}{k_\T{B} B \left(4\pi \sum_p T_\T{b}^p \bar{A}^p_\T{eff}(\omega_0)\lambda_p^{-2}\right)},
    \label{eq:snr-tv}
\end{equation}
which, in contrast to the LTI case, clearly depends on signal and noise characteristics in and out of the observation band.

\section{Summary of effective noise temperatures}
\label{sec:summary}

In the preceding sections, emphasis was placed on deriving expressions for the realized SNR of a time-varying system by extending its definition from the LTI case to account for out-of-band coupling.  In practice, the external SNR, not the external noise temperature, plays a role in the overall performance of a system.  However, since signal power is typically not under the control of receiver designers, here we summarize the effective noise temperatures of both the LTI and time-varying cases for both the isotropic cases considered in previous sections as well as arbitrary, non-isotropic external brightness temperatures.

Repeating the LTI analysis in Sec.~\ref{sec:lti} without the assumption of an isotropic noise temperature leads to the external noise temperature 
\begin{equation}
    T_\T{A}^\T{LTI} = \frac{1}{\lambda_0^2}\int A_\T{eff}(\omega_0,\kv) T_\T{b}(\omega_0,\kv)\T{d}\kv,
    \label{eq:result-lti-noniso}
\end{equation}
aligning with the methodology taken to derive existing results~\cite[\S 2.18]{balanis2016antenna}, though here the use of realized effective aperture allows for a compact and practical incorporation of efficiency and mismatch effects. 
 Simplification to the isotropic noise model in \eqref{eq:noise-model} leads to
\begin{equation}
    T_\T{A}^\T{LTI} = \eta\tau T_\T{b},
    \label{eq:result-lti-ta-iso}
\end{equation}
which appears in \eqref{eq:lti-noise}.

Similarly for a time-varying system, the non-isotropic effective noise temperature becomes
\begin{equation}
    T_\T{A}^\T{TV} = \sum_p \frac{1}{\lambda_p^2} \int A_\T{eff}^p(\omega_0,\kv) T_\T{b}^p(\kv)\T{d}\kv,
    \label{eq:result-tv-noniso}
\end{equation}
which simplifies to the isotropic case
\begin{equation}
    T_\T{A}^\T{TV} = 4\pi \sum_p \frac{T_\T{b}^p\bar{A}_\T{eff}^p(\omega_0)}{\lambda_p^2},
    \label{eq:result-tv-ta-iso}
\end{equation}
appearing in \eqref{eq:tv-littlepnoise}-\eqref{eq:snr-tv}.  The above expression constitutes the core theoretical development of the paper.  Throughout the following sections, we study the nature of the cross-frequency effective aperture terms $\bar{A}_\T{eff}^p$ and effective noise temperature $T_\T{A}^\T{TV}$ of several classes of receive systems.

\section{Demonstrative analysis}
\label{sec:demo}

To demonstrate the preceding analysis of effective noise temperatures without going into the design and optimization of a particular time-varying receiver, we opt to first study a simplified system modeled by only a few scalar parameters.
Consider an LTI small dipole antenna with maximum directivity $D = 1.5$.  For any mismatch-efficiency product $\eta\tau$, the associated realized effective aperture $A_\T{eff}$ is given by \eqref{eq:aeff-g}, though subsequent results are normalized to be independent of this product, so it can be chosen arbitrarily.  The average effective aperture $\bar{A}_\T{eff}$ is given by the right-hand-side of \eqref{eq:aeff-bar} multiplied by an additonal factor of $(4\pi)^{-1}$. The antenna is immersed in a noisy background characterized by the isotropic brightness temperature $T_\T{b}(\omega)$.  Based on the prescribed directivity and brightness temperature, the SNR of this receiver, normalized by $S_0/(k_\T{B}B)$, can be computed from~\eqref{eq:snr-lti}.

For comparison with the LTI case, a hypothetical modification to the receiving dipole is made using periodically time-varying loads.  We assume that the loads oscillate with fundamental frequency $\omega_\T{m} = \omega_0/2$ and that their introduction scales the parameters $A_\T{eff}^0$ and $\{\bar{A}_\T{eff}^p\}$ relative to the dipole's effective aperture $A_\T{eff}$ and average effective aperture $\bar{A}_\T{eff}$, respectively, with the form
\begin{equation}
    A_\T{eff}^0 = \alpha A_\T{eff},
    \label{eq:example-aeff0-1}
\end{equation}
\begin{equation}
    \bar{A}_\T{eff}^0 = \bar{A}_\T{eff},
    \label{eq:example-aeff0-2}
\end{equation}
and
\begin{equation}
    \bar{A}_\T{eff}^{\pm 1} = \beta\bar{A}_\T{eff}.
\end{equation}
All terms $\bar{A}_\T{eff}^p$ with $|p|>1$ are assumed to be zero.  The scaling factor $\alpha$ represents an increase in the signal frequency effective aperture (analogous to increased directivity) while the factor $\beta$ controls the relative magnitude of the first-harmonic cross-frequency effective aperture terms.  As in the LTI case, the SNR for the time-varying receiver can be calculated using \eqref{eq:snr-tv} and normalized by $S_0/(k_\T{B}B)$.  

The SNR performance of either LTI or time-varying systems clearly depends on the nature of the noise environment, characterized here by the brightness temperature $T_\T{b}$.  To demonstrate this dependence, we study the relative SNR behavior of both systems in two unique noise environments representing two practical scenarios.

\subsection{Flat noise temperature}

\begin{figure}
    \centering
    \includegraphics[width=3.2in]{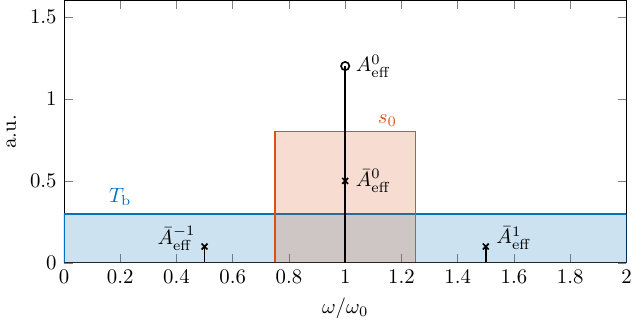}\\
    \includegraphics[width=3.5in]{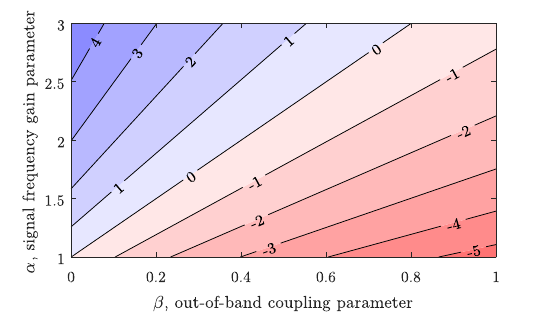}
    \caption{Schematic depiction of an example calculation using a flat external brightness temperature (top panel).  The incoming spectrum is divided into signal (blue) and noise (red).  Markers denote the terms $A_\T{eff}^0$ (circle) and $\bar{A}_\T{eff}^p$ ($\times$).  Resulting ratio of TV and LTI SNR levels, in dB (bottom panel).}
    \label{fig:ex-uniform}
\end{figure}

In the first example, we consider a brightness temperature that is constant over all frequencies, see Fig.~\ref{fig:ex-uniform}.  In this case the noise environment at the fundamental, upper first harmonic, and lower first harmonic frequencies are identical.  This is representative of an unoccupied band having relatively uniform noise characteristics.  

In the lower panel of Fig.~\ref{fig:ex-uniform}, we plot the ratio of SNRs realized by the LTI and time-varying systems.  Recall that higher values of the parameter $\alpha$ indicate effective aperture enhancement at the signal frequency, whereas higher values of the parameter $\beta$ indicate higher coupling from the two first harmonics into the observation bandwidth.  Intuitively, we see that higher gain parameters lead to improved SNR, however this effect is directly canceled by increasing out-of-band coupling.  For cases of high out-of-band-coupling, the external SNR of the time-varying receiver is lower than that of the LTI system, despite improved signal frequency effective aperture. Thus this simple example demonstrates that the effective aperture at the carrier frequency $A_\T{eff}^0$ alone is not sufficient to characterize the performance of an externally noise limited time-varying receiver.

\subsection{Noisy neighbors}

\begin{figure}
    \centering
    \includegraphics[width=3.2in]{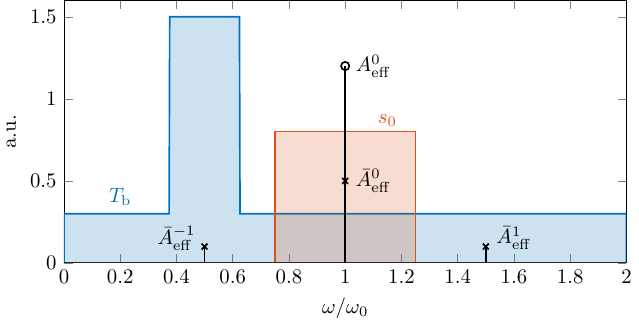}\\
    \includegraphics[width=3.5in]{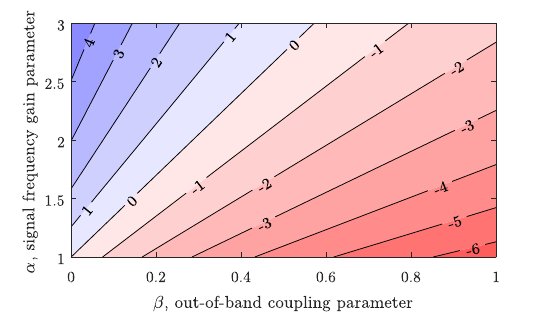}
    \caption{Schematic depiction of an example calculation using a non-uniform external brightness temperature (top panel).  The incoming spectrum is divided into signal (blue) and noise (red).  Markers denote the terms $A_\T{eff}^0$ (circle) and $\bar{A}_\T{eff}^p$ ($\times$).  Resulting ratio of TV and LTI SNR levels, in dB (bottom panel).}
    \label{fig:ex-noisy}
\end{figure}

In a more extreme example of detrimental out-of-band coupling, we consider the scenario depicted in the top panel of Fig.~\ref{fig:ex-noisy}, where the external brightness temperature is much higher over a frequency range including one or more of the modulation harmonics of the system.  This situation could be representative of a number of receiving systems.  For relatively small modulation frequencies $\omega_\T{m}\ll\omega_0$, the $|p|=1$ harmonics could coincide with neighboring occupied channels within the same communications protocol (e.g., neighboring WiFi channels).  In systems using high modulation rates $\omega_\T{m}\sim\omega_0$ to realize parametric amplification, this represents the coincidental alignment of a system's harmonics with a strong interfering signal occupying an entirely separate part of the spectrum (e.g., overlapping with an occupied ISM band, broadcast band, or naturally noisy part of the spectrum, such as HF).

The bottom panel of Fig.~\ref{fig:ex-noisy} shows the relative SNR performance of the LTI and time-varying receiver in this particular noise scenario.  We observe the same features as in the previous example, with increased penalization of the out-of-band coupling parameter due to the strong interferer coinciding with the first harmonic.  These results suggest that out-of-band coupling presents a larger detriment to systems with harmonics falling on occupied or exceptionally noisy bands.

\subsection{Discussion}
By the definition in \eqref{eq:example-aeff0-1}-\eqref{eq:example-aeff0-2}, the scaling parameter $\alpha$ governs directivity increases induced by some hypothetcal time-varying behavior, while the parameter $\beta$ represents a penalization by out-of-band coupling.  Intuitively, the preceding analyses demonstrate that increases in received signal (higher $\alpha$) can be negated by sufficiently high increases in out-of-band noise (higher $\beta$) and that this issue can be exacerbated in situations where harmonics of the time-varying system coincide with high-noise bands.

However, not all time-varying systems may be accurately represented by the toy problem defined in \eqref{eq:example-aeff0-1}-\eqref{eq:example-aeff0-2} and a more severe conclusion may be drawn by considering a small variation where both the average and peak effective apertures are modified by the parameter $\alpha$,
\begin{equation}
    A_\T{eff}^0 = \alpha A_\T{eff},
    \label{eq:example-aeff0-3}
\end{equation}
\begin{equation}
    \bar{A}_\T{eff}^0 = \alpha\bar{A}_\T{eff}
    \label{eq:example-aeff0-4}.
\end{equation}
In this case, the parameter $\alpha$ represents an increase in overall system gain while leaving directivity unchanged, \ie, amplification without changing the antenna's radiation pattern.  Like in the previous example, this situation leads inevitably to increased effective noise temperature \eqref{eq:result-tv-ta-iso} for any non-zero value of out-of-band coupling $\beta$, but also carries the additional consequence that, by \eqref{eq:snr-tv}, the received SNR must also lower as well.  This is a restatement of the fact that antenna gain does little to improve system performance in externally noise-limited systems.  However, for internally noise limited systems, an optimum configuration of system gain may exist which increases overall SNR despite increasing the external noise temperature through out-of-band coupling.  This is a complex system-specific tradeoff space that should be explored thoroughly in the analysis of any real time-varying system with both internal and external noise sources.

\section{Degenerate Mode Parametric Amplification}
\label{sec:pa}

To study the out-of-band coupling effects in a practical system, we consider the dynamically-tuned architecture described in~\cite{mekawy2021parametric}, where a small loop antenna is interfaced to a load impedance by a time-varying capacitor.  In~\cite{mekawy2021parametric}, the transmit properties of the system are considered, but here we study it as a receiver with no major changes to its design.  
The antenna itself is a planar, square perfectly conducting loop with outer dimension $\ell = 36~\T{mm}$ and trace width $w = 2~\T{mm}$.  The antenna was modeled using the method of moments (MoM) to calculate the output impedance $Z_\T{a}(\omega)$ and the open-circuit voltage $v_\T{oc}(\omega)$ present at the antenna terminals (modeled by a single RWG basis function edge~\cite{rao1982electromagnetic}) under  planewave illumination with electric field amplitude $1~\T{V}/\T{m}$ from various incident directions.  
The antenna is interfaced to a load impedance 
\begin{equation}
    Z_0 = rR_\T{a}(\omega_0)
\end{equation}
by a series time-varying capacitance
\begin{equation}
    C_\T{tv} = C_0\left[1+2M\cos(\omega_\T{m}t)\right],
\end{equation}
where $\omega_0$ denotes the intended design frequency of $300~\T{MHz}$.  The real-valued scaling factor $r$ determines the ratio of antenna resistance $R_\T{a}$ to load impedance $Z_0$ at the design frequency, while the parameter $M \in [0,1/2)$ is the capacitance modulation depth.  

Per~\cite{mekawy2021parametric}, the unmodulated capacitance $C_0$ is selected to resonate the antenna at the design frequency $\omega_0$.  To operate in degenerate mode, we set the pump frequency $\omega_\T{p} = 2\omega_0$.  The load and modulation parameters were swept to find a combination yielding moderate parametric gain, with values $r = 1.1$ and $M = 5\cdot10^{-4}$ used in all subsequent calculations.  The reference LTI system is defined by a standard conjugate match, \ie, $r = 1$, $M = 0$.

The constituent terms in the overall SNR of the time-varying and reference LTI systems are evaluated by two methods utilizing the equivalent Th\'evenin circuit model of the receiving antenna.  The first method uses a conversion matrix (CM) approach to time-varying circuit analysis applied to the simple circuit formed by the Th\'evenin antenna model and its load, functionally equivalent to a compressed CMMoM representation~\cite{bass2022conversion} adapted for computing received power, see Appendix~\ref{app:cmmom}.  The second approach is to calculate the steady-state response of the same system using a time-domain finite-difference (FD) simulation of the same circuit, see Appendix~\ref{app:tran} for implementation details.  

To facilitate replicating these calculations without the general CMMoM implementation described in Appendix A, we primarily report results obtained using a simplified antenna Th\'evenin equivalent circuit comprised of a resistance $R_\T{a} = 0.0523~\Omega$, inductance $L_\T{a} = 104.9~\T{nH}$, and an open circuit voltage~\cite{best2016optimizing}
\begin{equation}
    v_\T{oc} = E_0\frac{2c_0}{\omega}\sqrt{\frac{R_\T{a}D}{120}}
    \label{eq:pa-voc}
\end{equation}
where the directivity is modeled using the small antenna directivity approximation
\begin{equation}
    D(\theta) = \frac{3}{2}\sin^2\theta.
\end{equation}
With this approximation, the average effective aperture terms $\bar{A}_\T{eff}^p$ in \eqref{eq:aeff-bar} are equal to the maximum effective aperture multiplied by a factor of $2/3$, further reducing computational overhead for replicating the reported results.  Fullwave results using CMMoM are also reported for comparison and are discussed below.

\begin{figure}
    \centering
        \begin{circuitikz}[scale=0.8,transform shape]
        \draw (0,0) to[sV,l=$v_\T{oc}(t)$] (0,2) to[L,l_ = $L_\T{a}$] (2,2) to[R,l_ = $R_\T{a}$] (4,2) to[vC,l_ = $C(t)$] (6,2) to[short,i=$i$] (7,2) to[R,l_ = $R_\T{L}$] (7,0) to (0,0);
        \draw[dashed] (-1.6,-0.5) rectangle (4,3.2);
        \node at (1.2,2.75) {Antenna Th\'evenin Equivalent};

        \begin{scope}[shift={(3,5.5)},scale=0.1,rotate = 90]
        \def\w{2}
        \def\hlen{18}
        \draw[red] (-1.5*\w,-0.5*\w) rectangle (1.5*\w,1.5*\w);
        \draw[ultra thick,black](-\w,\w/2)--(\w,\w/2);
        \draw[fill=yellow!20] (-\w,0) -- (-\hlen,0) -- (-\hlen,2*\hlen)  -- (\hlen,2*\hlen) -- (\hlen,0) -- (\w,0) -- (\w,\w) -- (\hlen-\w,\w) -- (\hlen-\w,2*\hlen-\w)-- (-\hlen+\w,2*\hlen-\w) -- (-\hlen+\w,\w) -- (-\w,\w) -- (-\w,0);

        \draw[-stealth] (0,\hlen) -- (3*\w,\hlen) node[scale=10,right,rotate=-90] {$x$};
        \draw[-stealth] (0,\hlen) -- (0,\hlen+3*\w)node[scale=10,below,rotate=-90] {$y$}; 

        \draw[stealth-stealth] (-\hlen,2.2*\hlen) -- node[left,scale=10,rotate=-90] {$\ell$}(\hlen,2.2*\hlen);
        ;

        \draw[stealth-] (-\hlen,1.8*\hlen) -- (-1.1*\hlen,1.8*\hlen);
       \draw[stealth-] (-\hlen+\w,1.8*\hlen) -- (-0.9*\hlen+\w,1.8*\hlen) node[right,rotate=-90,scale=10] {$w$} ;
        \end{scope}

        \begin{scope}[shift={(5,3.85)},scale=0.8,transform shape]
        \draw[red] (-0.7,0) rectangle (2.5,4);
        \draw[fill=yellow!20] (0.2,0) -- (0.2,0.3) -- (1.8,0.3) -- (1.8,0);
        \draw[fill=yellow!20] (0.2,4) -- (0.2,3.7) -- (1.8,3.7) -- (1.8,4);
        \draw (1,0.3) to[R] (1,2) to[vC] (1,3.7);
        \node[scale=1.2] at (0.2,1.2) {$R_\T{L}$};
        \node[scale=1.2] at (0,2.8) {$C(t)$};
        \draw[dashed] (-0.7,4) -- (-2.3,2.5);
        \draw[dashed] (-0.7,0) -- (-2.3,1.5);
        \end{scope}
    \end{circuitikz}
    \caption{Parametrically loaded loop antenna and simplified equivalent circuit model. In simulation, the capacitor and resistor forming the antenna's load are modeled using lumped elements over a single RWG edge, drawn schematically in an exploded view in the geometry drawn above.}
    \label{fig:pa-circuit}
\end{figure}
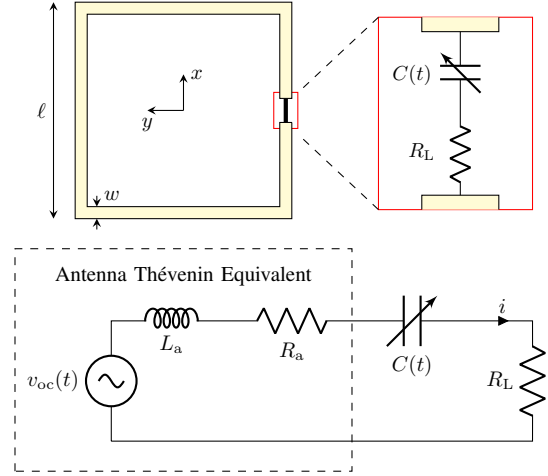

\begin{figure}
    \centering
    \includegraphics[width=3.5in]{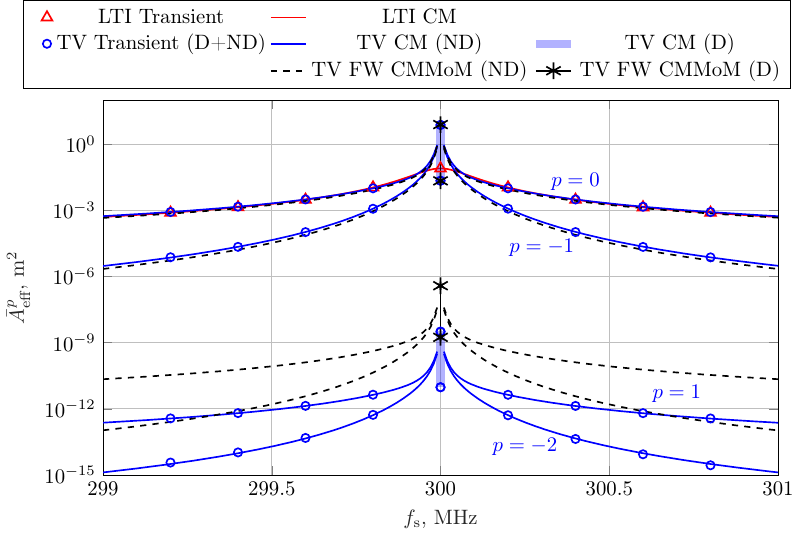}
    \caption{Average cross-frequency effective aperture terms $\bar{A}_\T{eff}^p$ for the parametric amplifier loop antenna depicted in Fig.~\ref{fig:pa-circuit}.  Results marked with ``FW'' are computed using an equivalent circuit based on fullwave CMMoM simulations of the loop structure.  All other results use the simplified circuit in Fig.~\ref{fig:pa-circuit}.  Time-varying (TV) results using CM or CMMoM are separated into degenerate (D) and non-degenerate (ND) cases.}
    \label{fig:loop-aeff}
\end{figure}

The frequency dependent average harmonic effective apertures $\bar{A}_\T{eff}^p(\omega)$ are computed by all three methods and plotted in Fig.~\ref{fig:loop-aeff}.   
At exactly 300 MHz, the time-varying antenna operates in a degenerate mode. The $p=0$ and $p=-1$ harmonic frequencies are equal and may constructively or destructively interfere at the output. Excitation by a pure tone at this frequency produces an effective aperture response that is dependent on the phase of the excitation relative to the pumping signal.  The range of values spanned at the degenerate operating point over all relative phases are delineated by the vertical light blue bars (simplified circuit model, CM), dual blue circular markers (simplified circuit model, transient method) and dual black markers (fullwave model, CMMoM). 

For observations made near the antenna's operating frequency of 300 MHz, the time-varying antenna's average effective aperture terms (blue traces and markers) for signals incident at the $p=0$ and $p=-1$ harmonic frequencies are greater than the single-frequency average effective aperture of the same antenna operated in an LTI mode (red traces and markers). Since these $\bar{A}_\T{eff}^0$ and $\bar{A}_\T{eff}^-1$ terms are of comparable magnitude and map the same neighborhood of frequencies into the output band, these terms are expected to dominate the time-varying antenna's effective external noise temperature.
Average effective aperture terms corresponding to the time-varying antenna's higher-order $p=-2$ and $p=1$ harmonics are several orders of magnitude lower, making it less likely that they will contribute appreciably to the effective external noise temperature.
Results for a general frequency-dependent model of the antenna's Th\'evenin equivalent circuit, obtained through the fullwave modeling described in Appendix A, are shown in black (CMMoM), with qualitatively similar behavior, though the $p = -2$ and $p=1$ harmonics show broader effective aperture bandwidth and overall higher values than in the simplified circuit model.

This parametric amplifying antenna system is an example of the situation described by \eqref{eq:example-aeff0-3} and \eqref{eq:example-aeff0-4}, in which amplification increases both the peak and average effective aperture values. In a spectrally flat noise environment, the effective antenna noise temperature is dominated by contributions from the $p=0$ and $p=-1$ harmonics. Since at the degeneracy point these are identical frequencies with equal values of $\bar{A}^p_\T{eff}$, noise incident in the observation band is effectively counted twice. The effective antenna temperature increases by a factor of two as a result of the additional noise power contribution from the $p=-1$ harmonic. However, spectrum mirroring in the $p=-1$ harmonic prevents an information-carrying desired signal from experiencing a similar bonus except under special circumstances \cite{blosser2024signal}.

\section{Time-modulated Arrays}
\label{sec:tma}
\begin{figure}
    \centering
        \begin{circuitikz}[scale=0.8,transform shape]
        \foreach \x in {0,1,2,3,7}
        {
        \draw (\x,0) to[switch] (\x,1);
        \draw[fill=black] (\x,1.3) circle (0.05);
        \node[antenna,scale=0.5,transform shape] at (\x,1) {};
        \draw[fill=white] (\x-0.4,-0.8) rectangle (\x+0.4,0.0);
        }
        \node at (5,0.5) {$\cdots$};
        
        \node at (-0.5,1.3) {$\rv_1$};
        \node at (-0.9,0.5) {$u_1(t)$};
        \node at (3.75,0.5) {$u_k(t)$};
        \node at (3.5,1.3) {$\rv_k$};
        \node at (7.75,0.5) {$u_K(t)$};
        \node at (7.5,1.3) {$\rv_K$};
        \node at (0,-0.4) {$A_1$};
        \node at (1,-0.4) {$A_2$};
        \node at (2,-0.4) {$A_3$};
        \node at (3,-0.4) {$A_k$};
        \node at (7,-0.4) {$A_K$};
        \draw (0,-0.8) -- (0,-1) -- (4,-1.6);
        \draw (1,-0.8) -- (1,-1) -- (4,-1.6);
        \draw (2,-0.8) -- (2,-1) -- (4,-1.6);
        \draw (3,-0.8) -- (3,-1) -- (4,-1.6);
        \draw (7,-0.8) -- (7,-1) -- (4,-1.6);
        \draw[fill=white] (4-0.4,-1.2) rectangle (4+0.4,-2) node[pos=0.5] {$\Sigma$};
        \draw[-stealth] (4,-2) -- (4,-2.3) node[below] {$v(t)$};

        \begin{scope}[shift={(5,3)},rotate=45]
        \draw (0.1,-0.7) -- (0.1,0.7);
        \draw (0.5,-0.7) -- (0.5,0.7) node[above left,rotate=-45] {$e(t-\rv\cdot\kv\omega/c)$};
        \draw (0.9,-0.7) -- (0.9,0.7);
        \draw[-stealth] (0.9,0) -- (-0.5,0) node[left,rotate=-45] {$\kv$};
        \end{scope}
    \end{circuitikz}
    \caption{Schematic of a simple ideal time-modulated array.  A set of $K$ antenna elements at locations $\rv_k$ are interfaced with a combiner $\Sigma$ with weights $A_k$ via independent switching networks characterized by the time-gating function $u_k(t)$.  The array is illuminated by uncorrelated plane waves from directions $\kv$ corresponding to either signals or noise.}
    \label{fig:tma-schem}
\end{figure}
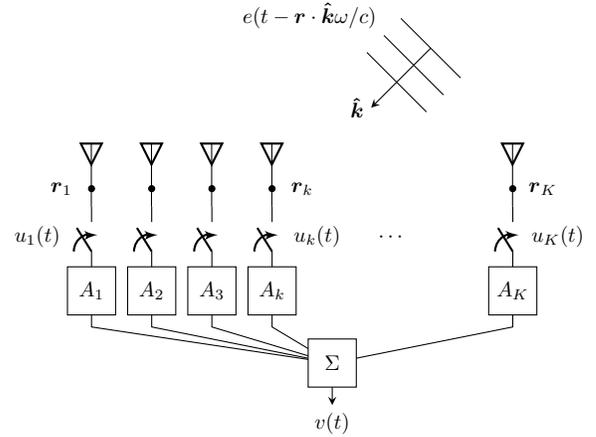

Another class of time-varying antennas are time-modulated arrays (TMA), where elements are interfaced with an array manifold via switching circuitry, sketched in Fig.~\ref{fig:tma-schem}.  By imposing periodic switching schemes, a variety of unique array behaviors can be realized, including sidelobe control and direction-dependent frequency conversion between bandwidths harmonically related to a nominal center frequency and the switching periodicity~\cite{kummer1963ultra,yang2004evaluation,poli2010pattern,maneiro2017time}.

A receiving TMA may convert signals between harmonically related frequencies, with the specific conversions corresponding to a variety of modes of operation.  For systems based on implementing spatial discrimination of signals sharing the same carrier frequency via conversion into harmonically-related bandwidths associated with unique TMA radiation patterns \cite[Fig.~1]{zhu2015signal}, a pre-switch bandpass filter is essential for reducing out-of-band coupling of interfers and noise into the observation bandwidth.  Here we apply the measures of effective external noise temperatures developed in previous sections to quantify the impact of this filtering stage in a simple, ideal TMA.

\subsection{TMA Model}
Consider the time-modulated array of isotropic elements with switch controls $u_k(t)$, effective lengths $\ell$, and locations $\V{r}_k$, as depicted in Fig.~\ref{fig:tma-schem}.  When illuminated with a plane wave with electric field $e$ incident from the direction $\kv$, the resulting voltage appearing at the output of the array manifold is
\begin{equation}
    v(t) = \ell\sum_k A_k u_k(t) e(t-\V{r}_k\cdot\kv\omega/c,\kv).
    \label{eq:tma-vk}
\end{equation}
Many manipulations, see App.~\ref{app:tma}, lead to the cross-frequency effective aperture for this system
\begin{equation}
    A^p_\T{eff}(\omega,\kv) = \frac{\eta_0\ell^2}{Z_0} \sum_{k,k'}  A_k A^*_{k'} U_k^p U_{k'}^{p*}\T{e}^{-\T{j}\phi^p_{kk'}},
    \label{eq:tma-aeff}
\end{equation}
where $U_k^p$ are the Fourier series coefficients representing $u_k(t)$ in harmonics of a fundamental modulation frequency $\omega_\T{m}$ and
\begin{equation}
    \phi_{kk'}^p = (\omega+p\omega_\T{m})(\kv\cdot(\rv_k-\rv_{k'}))/c.
    \label{eq:phikkdef}
\end{equation}  
Within the double sum found in \eqref{eq:tma-aeff}, conjugate symmetry under the exchange of the indices $k$ and $k'$ leads to purely real values of effective aperture.

For the particular case of sequential toggling of individual elements~\cite{yang2004evaluation}, the switch control mask takes the form
\begin{equation}
    u_k(t) = \begin{cases}
        0 & \text{if } t \in [0,t_k)\\
        1 & \text{if } t \in [t_k,t_k+\tau_k) \\
        0 & \text{if } t \in [t_k+\tau_k,T]
    \end{cases} ,
    \label{eq:tma-uk}
\end{equation}
which has Fourier coefficients
\begin{equation}
    U_k^p = \hat{\tau}_k \T{sinc} (\pi p \hat{\tau}_k)\T{e}^{-\T{j}\pi p (2\hat{t}_k+\hat{\tau}_k)}.
\end{equation}
Here $\hat{\tau}_k = \tau_k/T$ and $\hat{t}_k=t_k/T$ are normalized activation times, the sinc function is defined as $\T{sinc}\,x = \sin x / x$, and $T = 2\pi/\omega_\T{m}$ is the period of sequential switching.  

The form of \eqref{eq:tma-aeff} formalizes the point that a time-modulated array unintentionally maps harmonic inputs into an observation bandwidth.  Unlike the parametric amplification example in the previous section, however, the time-modulated array affords the possibility of bandpass filtering each antenna element prior to each switch, thus reducing the input signal at these unintended harmonics.  

\subsection{Analysis}
To study these effects in detail, we examine the behavior of cross-frequency effective aperture terms in \eqref{eq:tma-aeff} for a particular 8-element ideal linear time-modulated array described in \cite{zhu2015signal} with uniform half-wavelength spacing
\begin{equation}
    \V{r}_k = \u{x}k\lambda_0/2,
\end{equation}
uniform weights
\begin{equation}
    A_k = 1,
\end{equation}
two-cycle staggered element activation time
\begin{equation}
    \hat{t}_k = \text{mod}\left((k-1)/4,1\right),
\end{equation}
uniform activation periods
\begin{equation}
    \hat{\tau}_k = 1/4 \quad \forall k,
\end{equation}
and modulation rate related to operating frequency via a free parameter $M$, i.e.,
\begin{equation}
    T = 2\pi M/\omega_0.
\end{equation}

\begin{figure}
    \centering
    \includegraphics[width=3.5in]{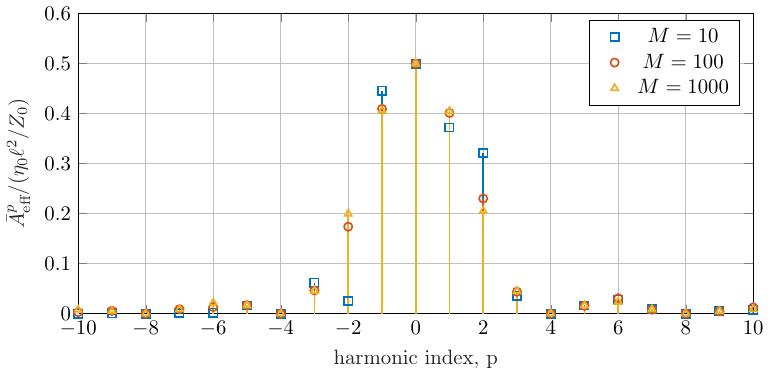}
    \caption{Cross-frequency effective aperture averages for an example 8-element time-modulated array.}
    \label{fig:tma-aeffpbar}
\end{figure}

The average harmonic effective apertures $\bar{A}_\T{eff}^p$ are computed for this model using \eqref{eq:tma-aeff} for varying relative modulation rates $M$ and plotted in Fig.~\ref{fig:tma-aeffpbar}. For validation, the same quantities are also computed directly in the time domain using \eqref{eq:tma-vk} with identical results, but those data are omitted for clarity.

As expected from \eqref{eq:tma-aeff}, results converge for low modulation rates (large $M$).  However, due to the imposed structure of two-cycle switching and thus \eqref{eq:tma-uk}, the limit of $M\rightarrow\infty$ does not correspond to the case of a static array, which would require $U_k^p = \delta_{p0}$, with $\delta_{ij}$ being the Kroenecker delta function.

Results in Fig.~\ref{fig:tma-aeffpbar}, demonstrate significant out-of-band coupling across all examples, particularly in the $p = \pm1$ harmonics.  The exact impact of this coupling on the total sytem external noise temperature \eqref{eq:result-tv-ta-iso} depends on the relative spacing of harmonic frequencies $\omega_0+p\omega_\T{m}$, their associated wavelengths $\lambda_p$, and the frequency dependent external noise profile $T_\T{b}^p$.  As a first-order approximation, however, we can examine the case of narrowband modulation where $\lambda_p\approx\lambda_0$ and the external noise environment is relatively constant, \ie, $T_\T{b}^p\approx T_\T{b}^0$.  Under these simplifications, and assuming a bandpass filtering stage that retains only harmonics $|p|\leq P$, the external noise temperature \eqref{eq:result-tv-ta-iso} reduces to a sum
\begin{equation}
    T_\T{A}^\T{TV} \approx \frac{4\pi T_\T{b}^0}{\lambda_0}\sum_{p=-P}^P \bar{A}_\T{eff}^p(\omega_0).
    \label{eq:tma-ta-simplified}
\end{equation}
Considering that the fundamental harmonic $p = 0$ must be kept by any filtering stages to capture the intended signal at frequency $\omega_0$, the relative increase in external noise temperatures can be studied by computing \eqref{eq:tma-ta-simplified} for varying filter widths $P\geq1$ relative to the limiting narrowband case of $P=0$.  The data in Fig.~\ref{fig:tma-aeffpbar} give a 4.18~dB and 5.35~dB increase in effective noise temperature for $P = 1$ and $P = 2$ filter widths, respectively, for the $M = 1000$ case. 

This example calculation contains many simplifications and approximations, but it demonstrates the important point that any time variation in a receiving system, including those commonly used in TMAs, inherently leads to coupling of out-of-band noise sources into the observation bandwidth.  Unlike previous examples of antennas with on-aperture loading in Secs.~\ref{sec:demo} and \ref{sec:pa}, however, the TMA architecture in Fig.~\ref{fig:tma-schem} affords the possibility of partially mitigating this adverse effect through the use of bandpass filtering stages located between the antenna and time-varying elements.  This bandpass filtering stage can reduce the incoming noise signals at out-of-band harmonics before they mix down into the observation bandwidth, with the overall effect of greatly reducing out-of-band effective aperture terms $\bar{A}_\T{eff}^p$.  This unique opportunity for exchanging bandwidth for favorable external noise temperature should be carefully considered in the design and analysis of TMAs in high-noise environments, particularly when modulation harmonics may lie in adjacent, high-interference bands, such as the example sketched in Fig.~\ref{fig:ex-noisy}.

\section{Conclusions}
The external noise power received by any antenna is quantified by the effective antenna noise temperature, which in this work is generalized to capture the intermodulation effects of time-varying receivers.   
Increases in antenna noise temperature depend directly on the average cross-frequency effective aperture at harmonic frequencies via \eqref{eq:result-tv-ta-iso}, particularly those with high external noise.  The examples presented here deal with a range of applications and system architectures, demonstrating the diversity of potential impacts on practical systems and laying a path for noise analysis in the design of time-varying receivers.  Future work on the optimization of modulation rates, filtering stages, and antenna architectures in time-varying systems may allow for the minimization (or maximization, if desired) of cross-frequency effective aperture terms as part of an overall design objective. In general, any gain, non-reciprocity, or other exotic behavior afforded by time-varying receiver systems should be carefully weighed against the additional external noise that is mixed into the observation bandwidth by time-varying devices.

\appendices

\section{Computation of cross-frequency effective aperture using conversion matrix method of moments}

\label{app:cmmom}
The effective aperture of an LTI antenna is straightforward to calculate using the method of moments~\cite[Ch. 10]{jin2015theory}.  Consider an antenna described by the impedance matrix $\M{Z}(\omega)$ which includes a terminating load impedance $R_0$ positioned on the port or basis function denoted by the index $\alpha$.  In this case, the effective aperture is given by
\begin{equation}
    A_\T{eff}(\omega,\kv) = \frac{R_\T{0}}{2}\left|\left[\M{Z}^{-1}\M{V}_{\kv}\right]_\alpha\right|^2
    \label{eq:aeffcmmom}
\end{equation}
where $\M{V}_{\kv}$ is an excitation vector associated with a plane wave carring unit power flux density at frequency $\omega$ from the direction $\kv$. 

The method of moments can be extended to periodically time-varying systems using a conversion matrix (CMMoM) approach~\cite{bass2022conversion}, which produces a conversion impedance matrix $\hat{\M{Z}}(\omega,\omega_\T{m})$ whose blocks represent coupling between harmonic frequencies $\omega+p\omega_\T{m}$ and $\omega+p'\omega_\T{m}$~\cite[\S 6.1.3]{maas2005noise}.  The elements of this impedance conversion matrix are defined as the ratio of voltage at basis function $\alpha$ and harmonic $p$ to the current at basis function $\beta$ and harmonic $p'$, \ie,
\begin{equation}
    Z_{\alpha\beta, pp'} = \frac{V_{\alpha,p}}{I_{\beta,p'}}.
\end{equation}
To compute the cross-frequency effective aperture, the system is illuminated by a plane wave with unit power flux density at the $p^\T{th}$ harmonic of the periodic system by constructing the excitation vector $\hat{\M{V}}^{\kv,p}$.  This vector is of full dimension corresponding to that of the impedance conversion matrix $\C{Z}$, with only the positions corresponding to the $p$ harmonic populated with values determined by the incident wave at that frequency.  All remaining entries are zero.  Driving the system with this excitation, the resulting current at the observation frequency $\omega_0$ over the termination resistor $R_0$ is sampled as before,
\begin{equation}
    A^p_\T{eff}(\omega,\kv) = \frac{R_\T{0}}{2}\left|\left[\hat{\M{Z}}^{-1}\hat{\M{V}}_{\kv,p}\right]_{\alpha,0}\right|^2,
    \label{eq:app-aeff-tv}
\end{equation}
where again dependence of both the excitation vector and conversion impedance matrix on $\omega$ and $\omega_m$ is suppressed.
With \eqref{eq:app-aeff-tv} implemented, integration over the far-field sphere can be carried out using an appropriate quadrature rule, such as Lebedev~\cite{LebedevLaikov_QuadratureRuleSphere}, to obtain average cross-frequency effective apertures.

The external noise temperature can also be written using the model's CMMoM representation via voltage and current noise covariance matrices \cite{russer2015}.  The general definitions of cross-frequency covariance matrices for voltages and currents ($\C{C}^V$ and $\C{C}^I$, respectively) on basis functions $\alpha,\beta$ for signals at frequencies $p,p'$ are
\begin{equation}
    C^V_{\alpha\beta,pp'} = \langle V_{\alpha,p}V^*_{\beta,p'}\rangle
\end{equation}
and
\begin{equation}
    C^I_{\alpha\beta,pp'} = \langle I_{\alpha,p}I^*_{\beta,p'}\rangle,
\end{equation}
where $\langle\cdot\rangle$ denotes the expected value. The utility of this representation lies in the fact that the total power spectral density dissipated at the $p=0$ harmonic in a termination $R_0$ located at the $\alpha$ basis function reads
\begin{equation}
    P = \frac{R_0\langle I_{\alpha,0}I^*_{\alpha,0}\rangle}{2} = \frac{R_0C^I_{\alpha\alpha,00}}{2},
\end{equation}
and when blocked in the same manner as the conversion impedance matrix $\C{Z}$, the covariance matrices are related~\cite{russer2015} via
\begin{equation}
    \C{C}^I = \C{Z}^{-1}\C{C}^V\C{Z}^{-\T{H}}.
    \label{eq:cv2ci}
\end{equation}

To compute the current covariance matrix (and 
 eventually the received power and noise temperature), we begin by following the assumption of \eqref{eq:n-distributed-source} that noise sources originating from different directions are uncorrelated. Additionally assuming the noise is wide-sense stationary, we compute the voltage covariance matrix for noise sources arising from a small portion of the far-field sphere $\T{d}\Omega$ centered on the direction~$\u{k}$,
 \begin{equation} \T{d}C^V_{\alpha\beta,pp'}(\u{k}) =  \delta_{pp'}V_{\kv,\alpha,p}V^{*}_{\kv,\beta,p'} \frac{2\eta_0k_\T{B}T_\T{b}^p}{\lambda_p^2} \T{d}\Omega,
\end{equation}
where the term $\delta_{pp'}$ arises because Fourier components of a wide-sense stationary random process at different frequencies are uncorrelated \cite[\S 2.4]{mandelWolf1995}.  As in \eqref{eq:prxn-tv-intermediate}, $\lambda_p$ and $T_\T{b}^p$ are the freespace wavelength and external brightness temperature at frequency $\omega+p\omega_\T{m}$.
Assuming an isotropic background and integrating over the far-field gives the complete matrix entry
\begin{equation}
C^V_{\alpha\beta,pp'} = \delta_{pp'} \frac{2\eta_0k_\T{B}T_\T{b}^p}{\lambda_p^2}\int_\Omega V_{\kv,\alpha,p}V^{*}_{\kv,\beta,p'}\T{d}\Omega
\end{equation}
Assembling all terms of this matrix and zero padding to match the dimension of the impedance conversion matrix, the voltage covariance matrix due to noise incident at the system's $p$th harmonic frequency can be written in the form 
\begin{equation}
    \C{C}^V_{p} = \frac{2\eta_0k_\T{B}T_\T{b}^p}{\lambda_p^2}\int_\Omega \C{V}_{\u{k},p}\C{V}^\T{H}_{\u{k},p}\T{d}\Omega.
    \label{eq:ccv-temp}
\end{equation}
The entire noise voltage covariance matrix $\C{C}^V$ is comprised of contributions from all harmonics $p$.  Summing these and applying \eqref{eq:cv2ci} leads to the current covariance matrix
\begin{equation}
    \C{C}^I = \sum_p\C{Z}^{-1}\C{C}^V_p\C{Z}^{-\T{H}}.
\end{equation}
Calculating the received power $P_0$ in the $p = 0$ harmonic leads to the effective external noise temperature
\begin{equation}
    T_\T{A}^\T{TV} = \frac{P_0}{k_\T{B}} = \frac{R_0}{2k_\T{B}} \left[\C{Z}^{-1}\C{C}^V\C{Z}^{-\T{H}}\right]_{\alpha\alpha,00}
\end{equation}
Combining the above expression with \eqref{eq:ccv-temp}, \eqref{eq:app-aeff-tv}, and \eqref{eq:aeffpbar-def} leads to \eqref{eq:result-tv-ta-iso}, demonstrating equivalency.  Hence the CMMoM framework for analyzing time-varying systems affords several avenues for the study of cross-frequency effective aperture and external noise temperature.

\section{Transient Simulation of Parametric Amplifier Example}

\label{app:tran}

Throughout this work, effective aperture quantities are primarily calculated in the frequency domain.  However, subtle difficulties arise in those calculations when the pumping frequency is selected such that any negative harmonic aligns with the negative image of the excitation frequency, i.e.,
\begin{equation}
    \omega_0 - p\omega_\T{m} = -\omega_0.
\end{equation}
To verify that these cases are treated correctly and to provide a ground truth comparison that is agnostic to frequency domain considerations, we treat the example discussed in Sec.~\ref{sec:pa} using a finite difference transient circuit simulation in addition to the CMMoM-based method presented in App.~\ref{app:cmmom}. In this transient computation, the terminated Th\'evenin equivalent circuit of the receiving antenna is driven with a continuous wave open-circuit voltage $v^\T{oc}(t)$ corresponding to a plane wave from a particular direction at each harmonic frequency $\omega_0+p\omega_\T{m}$ until steady state is reached. The Fourier transform of a windowed version of the steady state current response is then computed, from which the response at the observation frequency $\omega_0$ is extracted.  This amplitude is related to power dissipated in the load resistance normalized to  the incident power flux density to obtain the cross frequency effective aperture.  

This computation is straightforward to carry out in any transient circuit simulation tool (e.g., SPICE), provided a broadband circuit model of the recieving antenna and equivalent voltage source is available from some form of fullwave analysis.  For example, S-parameter blocks can be incorporated into convolution-based transient simulations in ADS~\cite{ADS} or explicit circuit models of broadband impedances may be synthesized using rational functions obtained via vector fitting~\cite{gustavsen1999rational,gustavsen2006improving} for use in SPICE-based circuit simulations.  The choice of method is also dependent on the availability of models for the time-varying components being studied, though simple versions of time-varying resistors, capacitors, and inductors can easily be modeled using multi-terminal nonlinear SPICE modules~\cite{Fisher1986Modeling}. 

Despite the generality (and potential implementation complexity) of the transient approach, the simplified RL model of the receiving loop antenna in Fig.~~\ref{fig:pa-circuit} 
 affords an extremely straightforward solution using a single-loop finite difference implementation.  Discretizing the solution on a grid of time points $t_n = n\Delta t$ and indexing all functions as $f(t_n) = f_n$, a marching-on in time equation for the current $i(t)$ flowing through the circuit is
\begin{equation}
    i_n = \left(v^\T{oc}_n + \frac{L_\T{a}}{\Delta t}i_{n-1} - \frac{\Delta t}{C_n}\sum_{n'=0}^{n-1}i_{n'}\right)\left(\frac{L_\T{a}}{\Delta t} + R\right)^{-1},
\end{equation}
where $R = R_\T{a} + R_\T{L}$ is the total loop impedance.  The frequency-depedent amplitude of the equivalent sinusoidal voltage source $v^\T{oc}(t)$ is given by \eqref{eq:pa-voc}.  All benchmark transient results in Fig.~\ref{fig:loop-aeff} were produced using this expression.

\section{Cross-frequency Effective Aperture of Time-modulated Arrays}
\label{app:tma}
Consider an ideal isotropic receiving element with effective length $\ell$ located at position $\rv_k$. If the element is illuminated by a plane wave incident from direction $\kv$ with amplitude spectral density $\tilde{e}(\omega,\kv)$ associated with a power spectral density per solid angle
\begin{equation}
    \tilde{y}(\omega,\kv) = \frac{|\tilde{e}(\omega,\kv)|^2}{2\eta_0},
\end{equation}
the voltage appearing over a load with impedance $Z_0$ is
\begin{equation}
    v_k(\omega,\kv) = \ell\T{e}^{-\T{j}\omega/c(\kv\cdot\rv_k)}\tilde{e}(\omega,\kv)
\end{equation}
If the load is instead connected to a switch characterized by the instantaneous transfer function $u_k(t)$, the voltage becomes
\begin{equation}
    v_k(\omega,\kv) = U_k(\omega)\star\left[\ell \T{e}^{-\T{j}\omega/c(\kv\cdot\rv_k)}\tilde{e}(\omega,\kv)\right],
\end{equation}
where $\star$ denotes convolution in the frequency domain.
Let the coefficients $A_k$ represent amplitude weights applied to the array elements.  The total array response is then
\begin{multline}
    v(\omega,\kv) =\\ \ell\sum_k A_k \int U_k(\omega-\omega')\T{e}^{-\T{j}\omega'/c(\kv\cdot\rv_k)}\tilde{e}(\omega',\kv)\T{d}\omega'.
\end{multline}
If the signals $\tilde{e}$ are random processes (\ie, noise signals), the expected value of the time-average power delivered to the load may be written
\begin{multline}
    \langle p(\omega,\kv) \rangle = \frac{\langle|v(\omega,\kv)|^2\rangle}{2Z_0} =\\ \frac{\ell^2}{2Z_0} \bigg\langle \bigg|\int \tilde{e}(\omega',\kv) \sum_k A_k  U_k(\omega-\omega')\T{e}^{-\T{j}\omega'/c(\kv\cdot\rv_k)}\T{d}\omega'\bigg|^2\bigg \rangle.
    \label{eq:TD-expected-power1}
\end{multline}
Expanding the squared magnitude term in \eqref{eq:TD-expected-power1} and combining terms gives
\begin{multline}
    \langle p(\omega,\kv) \rangle = \\ \frac{\ell^2}{2Z_0}  \int_{\omega'} \int_{\omega''} 
    \left\langle\tilde{e}(\omega',\kv)\tilde{e}^*(\omega'',\kv) \right\rangle \times \\
    \sum_{k,k'}  A_k A_{k'}^*  U_k(\omega-\omega') U_{k'}^*(\omega-\omega'') \times \\\T{e}^{-\T{j}\omega'/c(\kv\cdot\rv_k)} \T{e}^{\T{j}\omega''/c(\kv\cdot\rv_{k'})}\T{d}\omega'\T{d}\omega''.
\end{multline}
If the noise process is wide-sense stationary, the generalized Fourier components $\tilde{e}(\omega,\kv)$ are uncorrelated with respect to frequency \cite[\S 2.4]{mandelWolf1995} and their expected value collapses to $\langle|\tilde{e}(\omega',\kv)|^2\rangle \delta(\omega'-\omega'')$. Using this property, evaluation of one of the integrals produces
\begin{multline}
    \langle p(\omega,\kv) \rangle =  \frac{\ell^2}{2Z_0}  \int_{\omega'}  
    \left\langle\left|\tilde{e}(\omega',\kv)\right|^2 \right\rangle  \\
    \times\sum_{k,k'} A_k A_{k'}^*  U_k(\omega-\omega') U_{k'}^*(\omega-\omega') \\\times\T{e}^{-\T{j}\omega'(\kv\cdot(\rv_{k}-\rv_{k'}))/c}\ \T{d}\omega'.
\end{multline}
Assuming the incident power spectral density $\tilde{e}$ is also uncorrelated with respect to incidence angle, integrating over all incident angles to obtain the total noise power spectral density 
leads to another sum of uncorrelated random variables, and comparison with \eqref{eq:def-cross-aeff}
indicates that the cross-frequency effective aperture is given explicitly by
\begin{multline}
    a_\T{eff}(\omega,\omega',\kv) = \frac{\eta_0\ell^2}{Z_0}\times\\\sum_{k,k'}  A_k A_{k'}^*  U_k(\omega-\omega') U_{k'}^*(\omega-\omega')\T{e}^{-\T{j}\omega'(\kv\cdot(\rv_k-\rv_{k'}))/c}
\end{multline}
For periodic modulation, the control signals $U_k$ can be written in terms of a Fourier series with coefficients $U_k^p$, leading to the harmonic effective aperture in \eqref{eq:tma-aeff},
representing coupling of signals of frequency $\omega+p\omega_\T{m}$ incident from the direction $\kv$ into the observation frequency $\omega$ via \eqref{eq:aeffp-sum-form}.

\section*{Acknowledgment}
This research is based upon work supported in part by the Office of the Director of National Intelligence (ODNI), Intelligence Advanced Research Projects Activity (IARPA), via [2021-2106240007]. The views and conclusions contained herein are those of the authors and should not be interpreted as necessarily representing the official policies, either expressed or implied, of ODNI, IARPA, or the U.S. Government. The U.S. Government is authorized to reproduce and distribute reprints for governmental purposes notwithstanding any copyright annotation therein.

\bibliographystyle{ieeetr}
\bibliography{refs}

\end{document}